\documentclass[pra,twocolumn,groupedaddress]{revtex4-1}
\usepackage{graphicx}
\usepackage{graphics}

\def\ba{\begin{eqnarray}}
\def\ea{\end{eqnarray}}

\def\lb{\label}
\def\be{\begin{equation}}
\def\ee{\end{equation}}

\begin{document}

\author{G. Puentes and O. Santill\'an}
 \affiliation{Departamento de Fisica, Facultad de Ciencias Exactas y
Naturales, Pabell\'{o}n 1, Ciudad Universitaria, 1428 Buenos Aires,
Argentina}

\title{Zak Phase in Discrete-Time Quantum Walks}

\begin{abstract}
We report on a simple scheme that may present a non-trivial geometric Zak phase ($\Phi_{Zak}$) structure, which is based on discrete-time quantum walk architectures. By detecting the Zak phase difference between two trajectories connecting adjacent Dirac points where the quasi-energy gap closes for opposite values of quasi-momentum ($k$), it is possible to identify  geometric invariants. These geometric invariants correspond to $|\Phi_{Zak}^{+(-)}-\Phi_{Zak}^{-(+)}|=\pi$ and $|\Phi_{Zak}^{+(-)}-\Phi_{Zak}^{+(-)}|=0$, we argue that this effect can be directly measured. 
\end{abstract}
\date{\today}
\maketitle


\section{Introduction}

Quantum physics attaches a phase to particles due to the complex nature of the Hilbert space. Phases arising during the quantum evolution of a particle can have different origins. A type of geometric phase, the so-called Berry phase \cite{Berry}, can be ascribed to quantum particles which return adiabatically to their initial state, but remember the path they took by storing this information on a geometric phase ($\Phi$), defined as \cite{Berry, Hannay}:
\begin{equation}
e^{i\Phi}=\langle \psi_{\mathrm{ini}}|\psi_{\mathrm{final}} \rangle.
 \end{equation}

 Geometric phases carry a number of implications: they modify material properties of solids, such as conductivity in graphene \cite{Berrygraphene}, they are responsible for the emergence of surface edge-states in topological insulators, whose surface electrons experience a geometric phase \cite{Berrytopoinsul}, they can modify the outcome of molecular chemical reactions  \cite{Berrychemestry}, and  could even have implications for quantum information technology, via the  Majorana particle \cite{Berrymayorana}, or can bear close analogies to gauge field theories and differential geometry \cite{BerryGauge}. In this Letter, we present a simple system based on a discrete time quantum walk architecture, which presents a non-trival  Berry phase structure on the torus, i.e., the Zak phase \cite{Zak}. \\

~Discrete-time quantum walks (DTQWs) \cite{Aharonov} offer a versatile platform for the exploration of a wide range of non-trivial geometric and topological phenomena (experiment) \cite{Kitagawa, Crespi, Alberti}, and (theory) \cite{Kitagawa2,Obuse,Shikano2, Asboth,Wojcik,MoulierasJPB}. Further, QWs are roboust platforms for modelling a variety of dynamical processes from excitation transfer in spin chains \cite{Bose,Christandl} to energy transport in biological complexes \cite{Plenio}. They enable to study multi-path quantum inteference phenomena \cite{bosonsampling1,bosonsampling2,bosonsampling3,bosonsampling4}, and can provide for a route to validation of quantum complexity \cite{validation1,validation2}, and universal quantum computing \cite{Childs}. Moreover, multi-particle QWs warrant a powerfull tool for encoding information in an exponentially larger space, and for quantum simulations in biological, chemical and physical systems, in 1D and 2D geometries \cite{Peruzzo,
 Crespi,OBrien,Silberhorn2D}. \\

In this paper, we present a simple theoretical scheme for generation and detection of a non-trivial invariant geometric phase structure in 1D  DTQW architectures. The basic step in the standard DTQW is given by a unitary evolution operator $U(\theta)=TR_{\vec{n}}(\theta)$, where $R_{\vec{n}}(\theta)$ is a rotation along an arbitrary direction $\vec{n}=(n_{x},n_{y},n_{z})$, given by $$R_{\vec{n}}(\theta)=
\left( {\begin{array}{cc}
 \cos(\theta)-in_{z}\sin(\theta) & (in_{x}-n_{y})\sin(\theta)  \\
 (in_{x}+n_{y})\sin(\theta) & \cos(\theta) +in_{z}\sin(\theta)  \\
 \end{array} } \right), $$in the Pauli basis \cite{Pauli}. In this basis, the y-rotation is defined by a coin operator of the form  \cite{Pauli}.
$$R_{y}(\theta)=
\left( {\begin{array}{cc}
 \cos(\theta) & -\sin(\theta)  \\
 \sin(\theta) & \cos(\theta)  \\
 \end{array} } \right). $$  This is  
followed by a spin- or polarization-dependent translation $T$ given by 
$$
T=\sum_{x}|x+1\rangle\langle x | \otimes|H\rangle \langle H| +|x-1\rangle \langle x| \otimes |V\rangle \langle V|,
$$
 where $H=(1,0)^{T}$ and $V=(0,1)^{T}$.
The evolution operator for a discrete-time step is equivalent to that generated by a Hamiltonian $H(\theta)$, such that $U(\theta)=e^{-iH(\theta)}$ ($\hbar=1$), with $$H(\theta)=\int_{-\pi}^{\pi} dk[E_{\theta}(k)\vec{n}(k).\vec{\sigma}] \otimes |k \rangle \langle k|$$ and $\vec{\sigma}$ the Pauli matrices, which readily reveals the spin-orbit coupling mechanism in the system.~The quantum walk described by $U(\theta)$ has been realized experimentally in a number of systems \cite{photons,photons2,ions, coldatoms}, and has been shown to posses chiral symmetry, and display Dirac-like dispersion relation given by $\cos(E_{\theta}(k))=\cos(k)\cos(\theta)$. \\

Here, we present two different examples of non-trivial geometrical phase structure. 
The first DTQW protocol consists of two consecutvie spin-dpendent translations $T$ and rotations $R$, such that the unitary step becomes $U(\theta_1,\theta_2)=TR(\theta_1)TR(\theta_2)$, as described in detail in \cite{Kitagawa2}. The so-called ``split-step" quantum walk, has been shown to possess a non-trivial topological landscape characterized by topological sectors with different topological numbers, such as the winding number $W=0,1$. The dispersion relation for the split-step quantum walk results in \cite{Kitagawa2}:
$$
 \cos(E_{\theta,\phi}(k))=\cos(k)\cos(\theta_1)\cos(\theta_2)-\sin(\theta_1)\sin(\theta_2).
$$
The 3D-norm for decomposing the quantum walk Hamiltonian of the system in terms of Pauli matrices $H_{\mathrm{QW}}=E(k)\vec{n} \cdot \vec{\sigma}$  becomes \cite{Kitagawa}: \\
\begin{equation}
\begin{array}{ccc}
n_{\theta_1,\theta_2}^{x}(k)&=&\frac{\sin(k)\sin(\theta_1)\cos(\theta_2)}{\sin(E_{\theta_1,\theta_2}(k))}\\
n_{\theta_1,\theta_2}^{y}(k)&=&\frac{\cos(k)\sin(\theta_1)\cos(\theta_2)+\sin(\theta_2)\cos(\theta_1)}{\sin(E_{\theta_1,\theta_2}(k))}\\
n_{\theta_1,\theta_2}^{z}(k)&=&\frac{-\sin(k)\cos(\theta_2)\cos(\theta_1)}{\sin(E_{\theta_1,\theta_2}(k))}.\\
\end{array}
 \end{equation}
The dispersion relation and topological landscape for the split-step quantum walk was analyzed in detail in \cite{Kitagawa2}. We now turn to our second example. \\

The second example consists of two consecutive non-commuting rotations in the unitary step of the DTQW. The second rotation along the x-direction by an angle $\phi$, such that  the unitarity step becomes $U(\theta,\phi)=TR_{x}(\phi)R_{y}(\theta)$, where $R_{x}(\phi)$ is given in the same basis \cite{Pauli} by:
$$R_{x}(\phi)=
\left( {\begin{array}{cc}
 \cos(\phi) & i\sin(\phi)  \\
i \sin(\phi) & \cos(\phi)  \\
 \end{array} } \right).$$ The modified dispersion relation becomes:
\begin{equation}
 \cos(E_{\theta,\phi}(k))=\cos(k)\cos(\theta)\cos(\phi)+\sin(k)\sin(\theta)\sin(\phi),
\end{equation}
 where we recover the Dirac-like dispersion relation for $\phi=0$, as expected.
~The 3D-norm for decomposing the Hamiltonian of the system in terms of Pauli matrices  becomes: 
\begin{equation}
\begin{array}{ccc}
n_{\theta,\phi}^{x}(k)&=&\frac{-\cos(k)\sin(\phi)\cos(\theta)+\sin(k)\sin(\theta)\cos(\phi)}{\sin(E_{\theta,\phi}(k))}\\
n_{\theta,\phi}^{y}(k)&=&\frac{\cos(k)\sin(\theta)\cos(\phi)+\sin(k)\sin(\phi)\cos(\theta)}{\sin(E_{\theta,\phi}(k))}\\
n_{\theta,\phi}^{z}(k)&=&\frac{-\sin(k)\cos(\theta)\cos(\phi)+\cos(k)\sin(\theta)\sin(\phi)}{\sin(E_{\theta,\phi}(k))}.\\
\end{array}
 \end{equation}

\begin{figure} 
\includegraphics[width=0.9\linewidth]{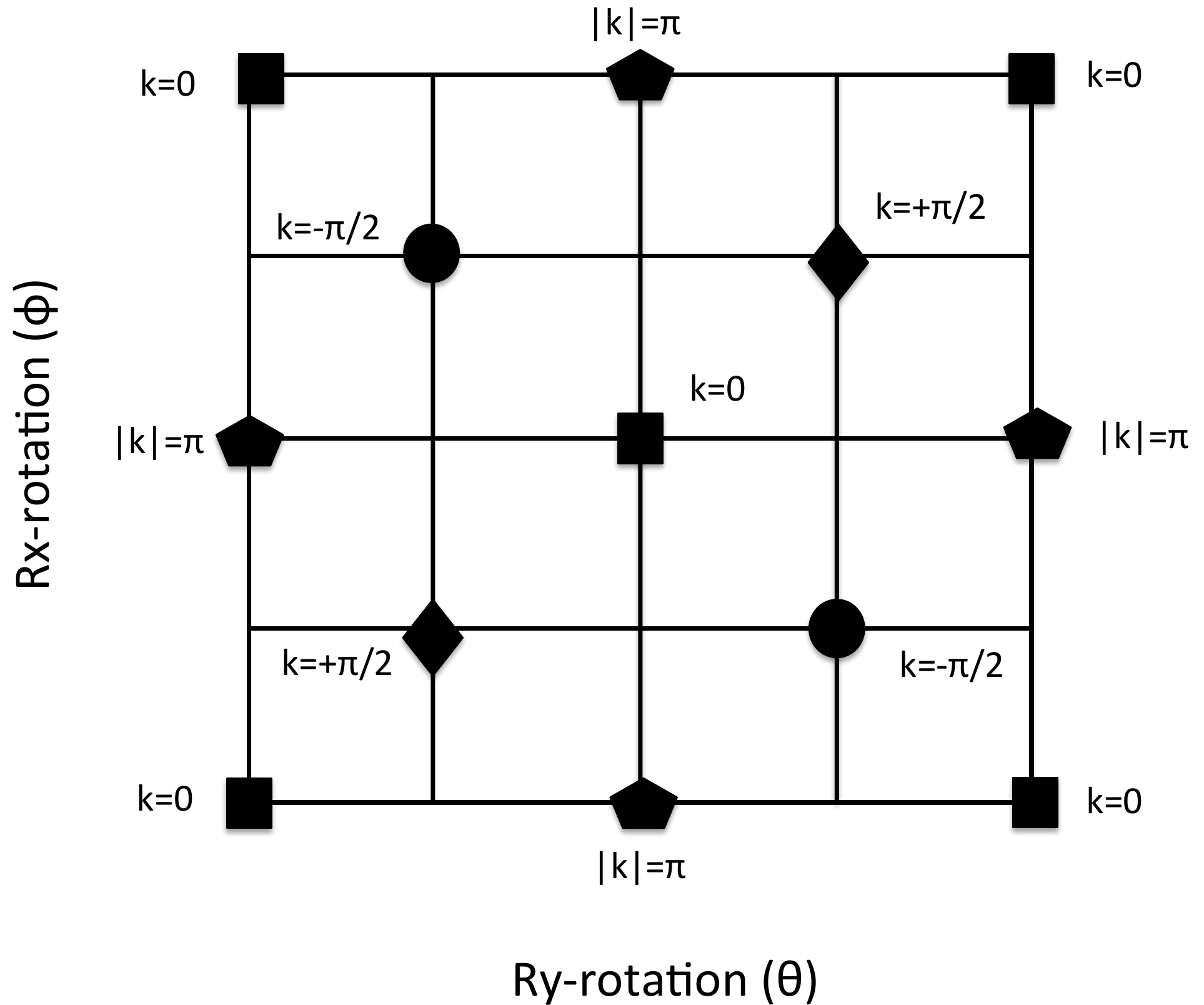} \caption{Non-trivial phase diagram for the quantum walk with consecutive non-commuting rotations, indicating gapless Dirac points where cuasi-energy gap closses for different values of quasi-momentum: Squares ($k=0$), pentagons ($|k|=\pi$), romboids ($k=+\pi/2$), circles ($k=-\pi/2$). }
\end{figure}

Dispersion relations for the DTQW with two consecutive non-commuting rotations within the first Brillouin zone are of the Dirac type (linear) at the set of gapless Dirac points, where the quasi-energy gap closes at $E(k)=0$. The second rotation enables to close the gap at zero energy for complementary points, and allows to create a non-trivial geometric phase structure in the system. In particular, this system has a non-trivial phase diagram with a larger number of gapless points for different momenta as compared to the system consisting of a single rotation. We calculated analitically the gapless Dirac points for the system. Using basic trigonometric considerations, it can be shown that the energy gap closes at 13 discrete points, for different values of quasi-momentum $k$. The phase diagram indicating the Dirac points where the gap closses for different momentum values is shown in Fig. 1. Squares correspond to Dirac points for $k=0$, circles correspond to Dirac points for $k=-\pi/2$, romboids correspond to Dirac points for $k=+\pi/2$, and pentagons correspond to Dirac points for $|k|=\pi$.  This geometric  structure in itself is novel and non-trivial and had not been studied in detail before. \\

\section{Zak Phase Calculation}

We will now give expressions for the Zak Phase in two different scenarios. These scenarios are casted by the following hamiltonian
\begin{equation}
H\sim n_x \sigma_x+n_y \sigma_y+ n_z \sigma_z,
\end{equation}
The hamiltonian to be described differ by a multiplying factor and by the expression of the $n_i$.
But since the eigenvectors are the only quantities of interest for the present problem, the overall constants of this Hamiltonian can be safely ignored.
Now, our generic hamiltonian is given by the matrix
\begin{equation}
H=
\left(
\begin{array}{cc}
  n_z \qquad    n_x-in_y\\
 n_x+i n_y \qquad   -n_z   
\end{array}
\right),
\ee
and has the following eigenvalues
\begin{equation}
\lambda=\pm \sqrt{n_x^2+n_y^2+n_z^2}
\end{equation}
The normalized eigenvectors then result
\begin{equation}
|V_\pm>= 
\left(
\begin{array}{cc}
  \frac{n_x+i n_y}{\sqrt{2n_x^2+ 2n_y^2+2n_z^2\mp 2n_z\sqrt{n_x^2+n_y^2+n_z^2}}}    \\
  \frac{n_z\mp \sqrt{n_x^2+n_y^2+n_z^2}}{\sqrt{2n_x^2+ 2n_y^2+2n_z^2\mp 2n_z\sqrt{n_x^2+n_y^2+n_z^2}}}   
\end{array}
\right)
\end{equation}

Note that the scaling $n_i\to\lambda n_i$ does not affect the result, as should be. This follows from the fact that two hamiltonians
related by a constant have the same eigenvectors.

The overall Zak phase for the problem is to be considered below is
\begin{equation}
Z=i\int_{-\pi/2}^{\pi/2} (<V_+|\partial_k V_+>+ <V_-|\partial_k V_->) dk.
\end{equation}
We will now apply these concepts to some specific examples.

\subsection{Split-step Quantum Walk}

We first consider the split-step quantum walk. This corresponds to a quantum walk with unitary step give by $U(\theta_1, \theta_2)=TR(\theta_1)TR(\theta_2)$, as proposed in \cite{Kitagawa2}. In this example the normasl $n_i$ are of the following form

\begin{equation}
\begin{array}{ccc}
n_{\theta_1,\theta_2}^{x}(k)&=&\frac{\sin(k)\sin(\theta_1)\cos(\theta_2)}{\sin(E_{\theta_1,\theta_2}(k))}\\
n_{\theta_1,\theta_2}^{y}(k)&=&\frac{\cos(k)\sin(\theta_1)\cos(\theta_2)+\sin(\theta_2)\cos(\theta_1)}{\sin(E_{\theta_1,\theta_2}(k))}\\
n_{\theta_1,\theta_2}^{z}(k)&=&\frac{-\sin(k)\cos(\theta_2)\cos(\theta_1)}{\sin(E_{\theta_1,\theta_2}(k))}.\\
\end{array}
 \end{equation}

We consider the particular case that $n_z=0$. By taking one of the angle parameters such that $n_z=0$, it follows that the eigenvectors of the Hamiltonian are:

\begin{equation}
|V_\pm>= 
\frac{1}{\sqrt{2}}\left(
\begin{array}{cc}
  e^{i\phi(k)}   \\
  \mp 1 
\end{array}
\right),\qquad 
\tan\phi(k)=\frac{n_y}{n_x}.
\end{equation}
There are two choices for $n_z=0$, which are $\theta_1=0$ or $\theta_2=0$.
The phase is in both cases results in
\begin{equation}
Z=Z_{+} +Z_{-}=i\int_{-\pi/2}^{\pi/2}dk <V_{+}|\partial_{k} V_{+}>
\end{equation}
\begin{equation}
+i\int_{-\pi/2}^{\pi/2} dk <V_-|\partial_k V_->=\phi(-\pi/2)-\phi(\pi/2),
\end{equation}
from where it follows that
\begin{equation}
Z=\frac{\tan(\theta_2)}{\tan(\theta_1)}.
\end{equation}
A plot of the Zak phase is presented in Fig. 2 (a).

\subsection{Quantum walk with non-commuting rotations}

The unitary step as described in the introduction results in $U(\theta,\phi)=TR_{x}(\phi)R_{y}(\theta)$.  The norms $n_i$ are of the following form
$$
n_x=-\cos(k) a+\sin(k)b,
 n_y=\cos(k) b+\sin(k) a,
$$
\be
m_z=\cos(k) c-\sin(k) d
\ee
with
$$
a=\sin(\phi)\cos(\theta)
$$
\be\lb{ang}
 b=\cos(\phi)\sin(\theta),
 \ee
$$
 c=\sin(\phi)\sin(\theta),\qquad d=\cos(\phi)\cos(\theta).
$$
the  angular functions defined above. The numerator $N_1$ is given by
\begin{equation}\lb{n1}
N_1=n_x+in_y=- \exp(-i k)(a-i b),
\end{equation}
The remaining numerator $N_2$ is
$$
N_2=n_z\mp \sqrt{n_x^2+n_y^2+n_z^2}=\cos(k) c-\sin(k) d
$$
\be\lb{n2}
\mp \sqrt{a^2+b^2+c^2 \cos^2(k)+d^2 \sin^2(k)-\sin(2k)cd}
\ee
On the other hand, the denominator  $D$ is reduced to 
$$
D_{\pm}=\sqrt{2n_x^2+ 2n_y^2+2n_z^2\mp 2n_z\sqrt{n_x^2+n_y^2+n_z^2}}
$$
$$
=\bigg(a^2+b^2+c^2 \cos^2(k)+d^2 \sin^2(k)-\sin(2k)cd
$$
$$
\mp (\cos(k) c-\sin(k) d)
$$
\be\lb{d}
\times \sqrt{a^2+b^2+c^2 \cos^2(k)+d^2 \sin^2(k)-\sin(2k)cd}\bigg)^{\frac{1}{2}}.
\ee
By taking into account these expressions, one finds that the eigenvectors may be expressed simply as
\begin{equation}
|V_\pm>= 
\left(
\begin{array}{cc}
 \frac{N_1}{D_\pm}    \\
  \frac{N_2}{D_\pm} 
\end{array}
\right),\qquad 
<V_\pm|=\bigg(\frac{N^\ast_1}{D_\pm},\; \frac{N_2}{D_\pm}\bigg)
\end{equation}
Then the calculation of the Zak phase
$$
Z=Z_+ +Z_-=i\int_{-\pi/2}^{\pi/2}dk <V_+|\partial_k V_+>
$$
\begin{equation}
+i\int_{-\pi/2}^{\pi/2} dk <V_-|\partial_k V_->
\end{equation}
requires to know the following quantities
$$
Z_\pm=i\int \bigg(\frac{N_1^\ast}{D_\pm^2}\partial_k N_1+\frac{N_2}{D_\pm^2}\partial_k N_2
$$
\begin{equation}
-\frac{(|N_1|^2+|N_2|^2)}{D_\pm^3}\partial_k D_\pm\bigg) dk,
\end{equation}
This expression can be simplified further. Due to  (\ref{n1}) it follows that the first term is real.  However, an inspection of (\ref{n2}) shows that the last two terms are purely imaginary. 
Since the overall phase should be real, it follows that these terms should cancel. This can be seen by taking into account that:
\be\lb{fb}
D_\pm=\sqrt{|N_1|^2+|N_2|^2},\qquad \partial_k |N_1|^2=0,
\ee
together with the fact that $N_2$ is real. Then  
\begin{equation}
\partial_k D_\pm=\frac{2N_2\partial_k N_2}{2D_\pm},
\end{equation}
where (\ref{fb}) has been taken into account. Therefore
$$
Z_\pm=i\int \bigg(\frac{N_1^\ast}{D_\pm^2}\partial_k N_1+\frac{N_2}{D_\pm^2}\partial_k N_2
$$
\begin{equation}
-\frac{(|N_1|^2+|N_2|^2)}{D_\pm^4}N_2\partial_k N_2\bigg) dk,
\end{equation}
but since  $D_\pm^2=|N_1|^2+|N_2|^2$ a simple calculation shows that the last two terms cancel each other. Thus the phase is
\begin{equation}
Z_\pm=i\int \frac{N_1^\ast\partial_k N_1}{D_\pm^2}dk.
\end{equation}
By taking into account  (\ref{d}) the phases are expressed as
\begin{equation}
Z_\pm=\int \frac{|N_1|^2}{D_\pm ^2}dk=\int_0^\pi \frac{(a^2+b^2)dk}{D_\pm^2} 
\end{equation}
 
We note that in this example the case $n_{z}=0$ is completely different than in the previous case, as it returns a trivial Zak phase $Z=\pi$, since the k-dependence vanishes. We note that for this system the Zak phase landscape can be obtained by numerical integration. In particular, at the Diract points indicated in Figure 1, the Zak phase is not defined.\\

\begin{figure} [t!]
\label{fig:2}
\hspace{-1cm}
\includegraphics[width=1.1\linewidth]{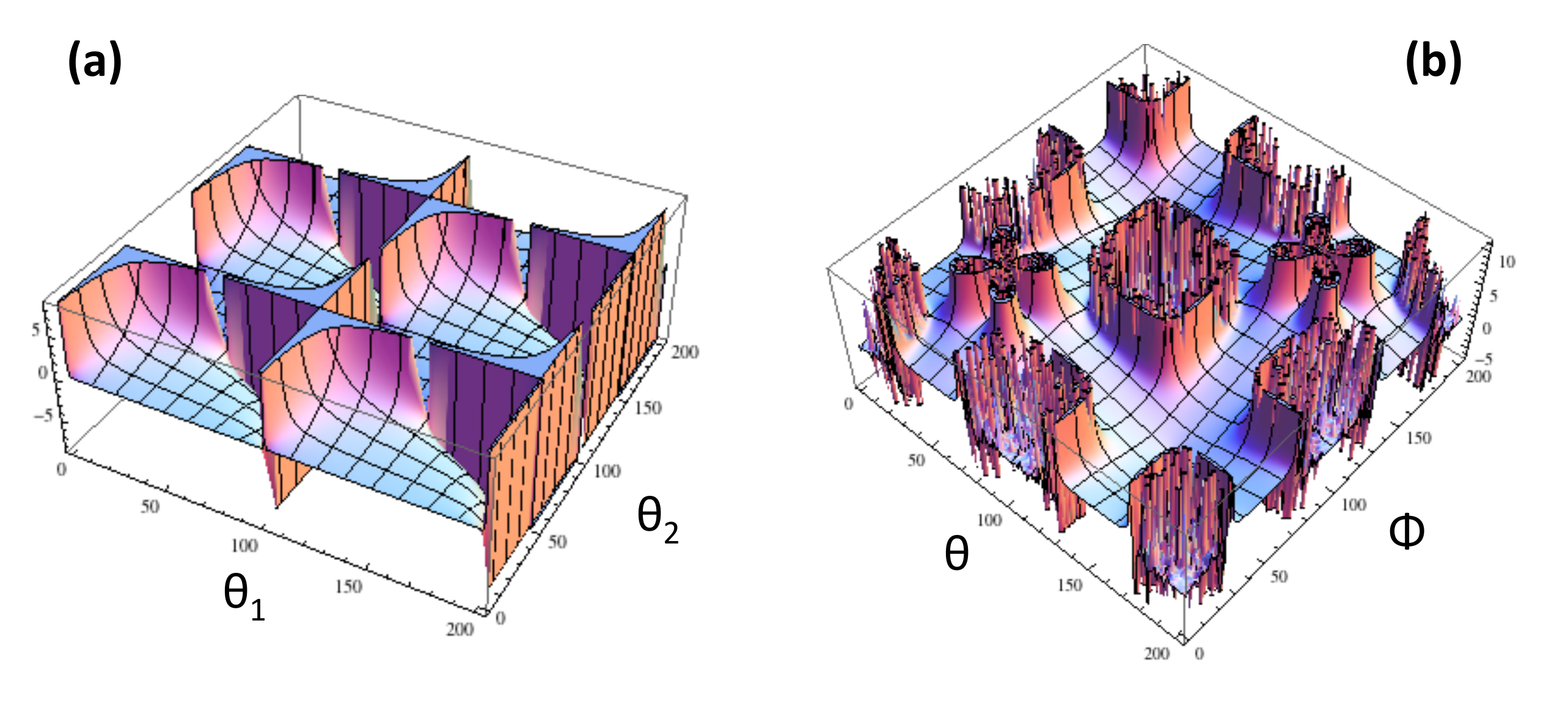} \caption{(a) Non-trivial geometric Zak phase landscape for "split-step" quantum walk, obtained analytically. (b) Non-trivial geometric Zak phase landscape for the quantum walk with non-commuting rotations, obtained by numeric integration.}
\end{figure}

\noindent A plot of the Zak phase $\Phi_{Zak}$ is shown in Fig. 2, for parameter values $\theta_{1,2}=[-\pi, \pi]$, and $\phi=[-\pi, \pi]$. (a) Zak phase for split-step quantum walk, given by the analytic expression $Z=\frac{\tan(\theta_2)}{\tan(\theta_1)}$, (b) Zak phase for quantum walk with non-commuting rotation obtained by numerical integration of  expression Eq. 2.30.\\

\section{Geometric Invariants}

\noindent It is well known that the Zak phase is not a geometric invariant, since it depends on the choice of origin of the Brillouin zone. However, a geometric invariant can be defined in terms of the Zak phase difference, between two evolved states ($|\psi^{-}_{\mathrm{final}} \rangle, |\psi^{+}_{\mathrm{final}} \rangle$) which differ on a geometric overall phase only, starting from the same initial state ($|\psi_{\mathrm{ini}}\rangle$) at a saddle Zak point. In this case, the Zak phase difference can be written as $\langle  \psi^{-}_{\mathrm{final}} |\psi^{+}_{\mathrm{final}}  \rangle=e^{i|\Phi_{Zak}^{+} - \Phi_{Zak}^{-}|}$. In \cite{ZakDemler, ZakLonghi}, experimental schemes for measuring the Zak phase difference for Bloch bands with two different dimerizations $D_{1,2}$, characterized by $\Delta k_{1}>0$ and $\Delta k_{2}<0$ where proposed, for \cite{ZakDemler} ultra-cold atoms, \cite{ZakLonghi} photons in waveguides. In our system, we can define a geometric analogue of
  dimerization, given by two evolved trajectories ($|\psi^{\pm}_{\mathrm{final}} \rangle$) with final parameter values associated with Dirac gapless points characterized by values of quasi-momentum $k$ of opposite sign, resulting in a relative quasi-momentum differences $\Delta k$ of opposite signs. These trajectories can be defined for all 13 Dirac points in the system since they are all analogue, up to arbitrary offsets which do not affect the calculation of the Zak phase difference. The two types of trajectories ($|\psi^{\pm}_{\mathrm{final}} \rangle$) are equivalent, up to an overall constant geometric phase. The appearance of a Zak phase difference of $\pi$ between the positive ($\Delta k>0$) and negative ($\Delta k<0$) branches, connecting singular points where the quasi-energy gap closes, signals a non-trivial geometric invariant structure in the system. On the other hand, for the case of trajectories connecting Dirac points between two positive ($\Delta k>0$) or two n
 egative ($\Delta k<0$) branches, the Zak phase difference should be zero. We note that in principle the dynamic phase does not cancel for each trajectory, but it does cancel for the difference between trajectories, as explained bellow. \\

\begin{figure} [b!]
\label{fig:2}
\includegraphics[width=0.8\linewidth]{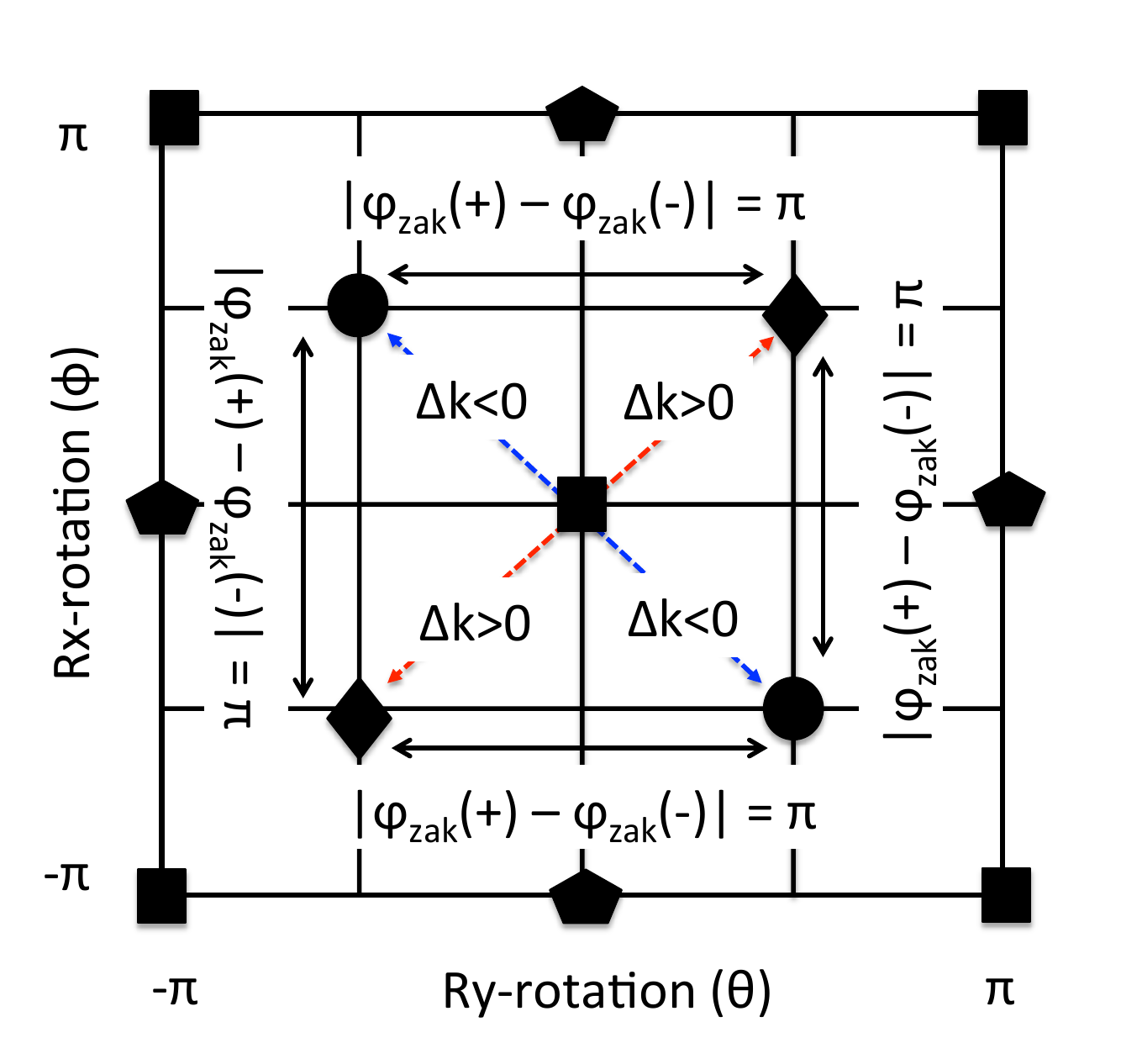} \caption{Geometric Zak phase difference $|\Phi_{Zak}^{\pm}-\Phi_{Zak}^{\pm}|$ for trajectories connecting Dirac points with positive(negative) momentum difference $\Delta k>(<)0$ (positive(negative) branch), indicated by red(blue) dotted lines. The Zak phase difference is expected to be of the form $|\Phi_{Zak}^{+(-)} - \Phi_{Zak}^{-(+)}|=\pi$ between trajectories connecting points with opposite $\Delta k$ and $|\Phi_{Zak}^{+(-)} -\Phi_{Zak}^{+(-)}|=0$ for trajectories with $\Delta k$ of the same sign (see text for details). Analogue geometric phase invariants can be defined for all other 12 singular points in the system where the quasi-energy gap closes. }
\end{figure}

\noindent The fact that the system aquires a different phase for the two trajectories ($|\psi^{\pm}_{\mathrm{final}} \rangle$) corresponding to Dirac points for quasi-momenta of opposite sign can be readily seen by replacing the final parameter values associated with each Dirac and Zak saddle point in the DTQW unitary step.  Let us consider the initial state $|\psi_{\mathrm{ini}}\rangle$ and the starting point for a trajectory at the center of phase diagram $(\theta=0, \phi=0)$. We note that the exact initial state and trajectory is irrelevant, as long as the final states are equivalent up to an overall phase, since the Zak phase difference for our system only depends on the final states $\langle  \psi^{-}_{\mathrm{final}} |\psi^{+}_{\mathrm{final}}  \rangle=e^{i|\Phi_{Zak}^{+} - \Phi_{Zak}^{-}|}$. We evolve the system to either of the four adjacent  Dirac points characterized by parameters $(\theta=\pi/2, \phi=\pi/2)$, $(\theta=\pi/2, \phi=-\pi/2)$, $(\theta=-\pi/2, \phi=\pi
 /2)$, $(\theta=-\pi/2, \phi=-\pi/2)$ by replacing these parameter values in the unitary step $U(\theta,\phi)=TR_{x}(\phi)R_{y}(\theta)$. For $\theta=\phi$ (positive branch) we find two  Dirac points, which corresponds to $\Delta k>0$, as indicated with romboid in Fig. 3. Evaluation of the product of  rotations $R_{x}(\phi)R_{y}(\theta)$, for these parameter values,  in a matrix which is proportional to $\sigma_{z}$ up to an overall phase $e^{i\pi/2N}$ dependent on the step number $N$. This is the dynamical phase acquired by the system during its evolution. Therefore, up to an overall dynamical phase, the system returns to its initial state in the coin (or spin) subspace, in addition to a translation in position to the site located at $x=+(-)N$ for initial state V(H) polarized $(0(1),1(0))^{T}$, due to the sequential translation operation $T^N$. For the case of the trajectories evolved to adjacent Dirac points characterized by $\theta= -\phi$ (negative branch),  other two Dirac points are identified corresponding to $\Delta k<0$, as indicated with circles in Fig. 3. The composed rotation for these parameters becomes $$\left( {\begin{array}{cc}
-i & 0  \\
0 & i  \\
 \end{array} } \right)=e^{i\pi}\left( {\begin{array}{cc}
i & 0  \\
0 & -i  \\
 \end{array} } \right).$$This corresponds to exactly the same evolution with the same dynamical phase dependent on the number of steps $e^{iN\pi/2}$, but multiplied by an overall constant geometric phase $e^{i\pi}$. Therefore, in this case the system is not only displaced by the translation operator $T^N$ to positions $x=\pm N$, but in addition it aquires an overall constant geometric Zak phase of $\pi$, as well as an overall dynamic phase which cancels out when measuring the phase difference $|\Phi_{Zak}^{+(-)}- \Phi_{Zak}^{-(+)}|$. Measurement of a Zak phase difference of either $\pi$ or $0$ between trajectories ($|\psi^{\pm}_{\mathrm{final}} \rangle$) corresponding to adjacent Dirac points would prove a non-trivial invariant geometric pattern in the system, which repeats itself over the entire Brillouin zone.  \\

Numerical simulations of the evolution of the  probability distribution  are shown in Figure 4, The overall phase (both dynamic and geometric) is of course not apparent in the probability distribution. The actual initial state and trajectory are of no concern, since the Zak phase difference only depends on the final states via $$\langle  \psi^{-}_{\mathrm{final}} |\psi^{+}_{\mathrm{final}}  \rangle=e^{i|\Phi_{Zak}^{+} - \Phi_{Zak}^{-}|}$$. Fig. 4 (a) is the probability distribution for the initial state $|\psi_{\mathrm{ini}}\rangle$ (step $N=0$) located at $x=0$, with righ-hand circular polarization $1/\sqrt{2}(1,i)^{T}$, Fig. 4 (b) is the probability distribution for an evolved state $|\psi_{\mathrm{final}} \rangle$ with arbitrary parameters $(\theta= \pi/4, \phi=0)$ after $N=7$ steps, which displays the typical ballistic delocalization associated with the Hadamard DTQW, Fig. 4(c) is the probability for the evolved state $|\psi^{-}_{\mathrm{final}} \rangle$ at the adjacent Dirac point located at $(\theta= \pi/2, \phi=-\pi/2)$ (negative branch) after $N=7$ steps, where the system aquires an overall dynamic phase $e^{iN\pi/2}$ and a constant geometric Zak phase $e^{i\pi}$ as explained above, and Fig. 4 (d) is the probability for the evolved state $|\psi^{+}_{\mathrm{final}} \rangle$ at another adjacent Dirac point located at $(\theta= \pi/2, \phi=\pi/2)$ (positive branch) after $N=7$ steps where the system acquires only an overall dynamical phase $e^{iN\pi/2}$ but no  geometric Zak phase, according to our definition of the Brillouin zone. We note that the absolute Zak phase aquired in each trajectory is somewhat arbitrary and depends on the choice of origin, therefore it is not invariant. On the other hand, the phase difference picked up by the two trajectories described in Fig. 4 (c) and 4 (d) is geometry invariant and can be directly measured, this  corresponds to a Zak phase difference $|\Phi_{Zak}^{+} -\Phi_{Zak}^{-}|=\pi$. \\

\begin{figure} [h!]
\label{fig:2}
\hspace{-0.4cm}
\includegraphics[width=1\linewidth]{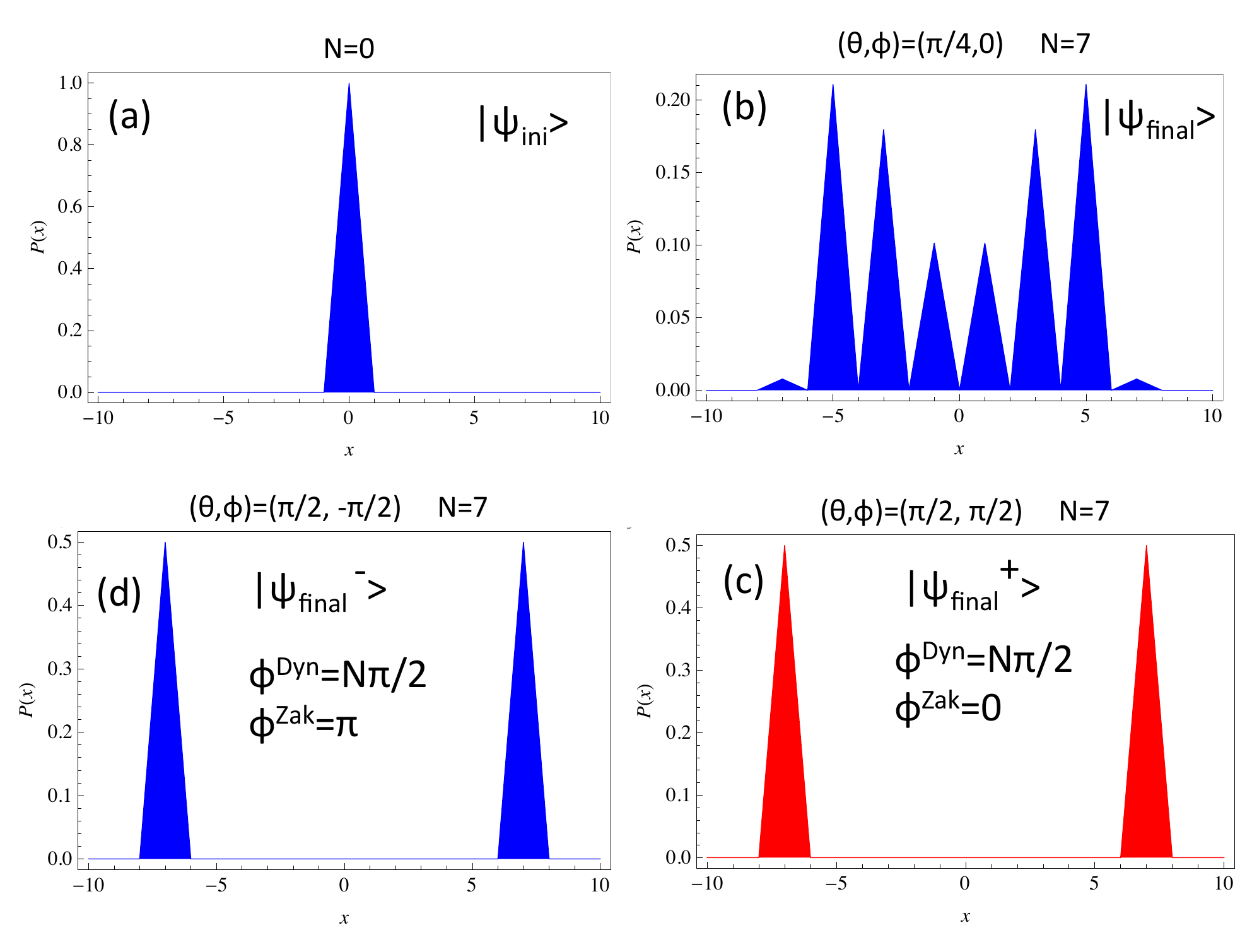} \caption{(a) Initial state ($N=0$) located at $x=0$, with  polarization $1/\sqrt{2}(1,i)^{T}$, (b) probability distribution after $N=7$ steps with parameters $(\theta= \pi/4, \phi=0)$ corresponding to Hadamard qunatum walk, (c) probability distribution after $N=7$ steps with parameters $(\theta= \pi/2, \phi=-\pi/2)$. With probability $P=1/2$ the system is translated to the left at position $x=-N$ and with probability $P=1/2$ the system is translated to the right position $x=N$. During this evolution the system aquires a dynamical phase $e^{iN\pi/2}$, and a non-zero geometric Zak phase $e^{i\pi}$, (d) probability distribution after $N=7$ steps for parameters $(\theta= \pi/2, \phi=\pi/2)$. With probability $P=1/2$ the system is translated to the left at position $x=-N$ and with probability $P=1/2$ the system is translated to the right at position $x=N$. During this evolution the system acquires a dynamical phase $e^{iN\pi/2}$ and 
 zero geometric Zak phase. Note that the absolute Zak phase for each trajectory depends on the definition of the Brillouin zone. The Zak phase difference between these two trajectories $|\Phi_{Zak}^{+(-)} - \Phi_{Zak}^{-(+)}|=\pi$ is a geometric invariant which can be measured. Equivalent geometric invariants can be measured for all the 13 Dirac points in the system. }
\end{figure}

\section{Discussion}

\noindent A simple experimental scheme to measure the Zak phase difference can be envisioned. Starting from any Dirac point in the parameter space, the system can be forced to evolve toward adjacent Dirac points through two different paths, corresponding to the positive ($\Delta k>0$) branch and the negative ($\Delta k<0$) branch. By shifting the origin of the Brillouin zone to the chosen Dirac point, the final parameter values for the negative path are characterized by $(\theta=+(-)\pi/2, \phi=-(+)\pi/2)$, while the final parameter values for the positive path correspond to  $(\theta=+(-)\pi/2, \phi=+(-)\pi/2)$. For each path, a different geometric  Zak phase should be aquired, such that $|\Phi_{Zak}^{+(-)}-\Phi_{Zak}^{-(+)}|=\pi$. This phase difference can be measured by recombining the two evolved states, in the case of photons by interferring the two evolved states via a Mach-Zehnder interferometer. On the other hand, for the case of two trajectories evolving along two positive or negative branches, the Zak phase difference should be zero $|\Phi_{Zak}^{+(-)}-\Phi_{Zak}^{+(-)}|=0$. A suitable scheme for detection of the Zak phase difference in a photonic system is described in \cite{HolonomicWhite}. 
\\

{\bf Acknowledgements:}  G. P. gratefully acknowledges financial support from PICT2014-1543 grant and Raices programme.

\end{document}